\documentstyle[psfig]{l-aa}
\newcommand{\etal}{et al. }
\begin{document}
\thesaurus{3
           (11.07.1;
	   11.09.4;
           11.19.2;
           13.09.1;
           13.09.3)}
\title{Mid--IR emission of galaxies in the Virgo cluster: II. Integrated properties.\thanks{
Based on observations with ISO, an ESA project with instruments founded by
ESA member states (especially the PI countries: France, Germany, the 
Netherlands and the United Kingdom) and with participation of ISAS and NASA}}
\label{}
\author{A. Boselli\inst{1}\and
        J. Lequeux\inst{2}\and
        M. Sauvage\inst{3}\and
        O. Boulade\inst{3}\and
        F. Boulanger\inst{4}\and
        D. Cesarsky\inst{4}\and
        C. Dupraz\inst{5}\and
        S. Madden\inst{3}\and
        F. Viallefond\inst{2}\and
        L. Vigroux\inst{3}}
\offprints{A. Boselli
           boselli@astrsp-mrs.fr}
\institute{Laboratoire d'Astronomie Spatiale, Traverse du Siphon, BP 8, 
           F-13376 Marseille CEDEX 12, France
\and
           Observatoire de Paris, 61 Av. de l'Observatoire, F-75014 Paris, 
           France
\and
           Service d'Astrophysique, Centre d'Etudes de Saclay, F-91191
           Gif sur Yvette, France
\and
           Institut d'Astrophysique Spatiale, Bat. 121, Universit\'e 
           Paris-Sud, F-91405 Orsay, France
\and
           Radioastronomie, Ecole Normale Sup\'erieure, 24 Rue Lhomond,
           F-75231 Paris CEDEX 05, France}
\date{Received , accepted  }
\maketitle        

\begin{abstract}

We analyse the integrated properties of the Mid-IR emission of a complete, optically selected
sample of galaxies in the Virgo cluster observed with the ISOCAM instrument on board the ISO 
satellite. The ISOCAM data allows us to construct the luminosity
distribution at 6.75 and 15 $\mu$m of galaxies for different morphological classes. These data
are used to study the spectral energy distribution of
galaxies of different type and luminosity in the wavelength range 2000 \AA - 100 $\mu$m.
The analysis shows that the Mid-IR emission up to 15 $\mu$m of optically-selected, normal 
early-type galaxies (E, S0 and S0a) is dominated by the Rayleigh-Jeans tail of the cold stellar 
component. The Mid-IR emission of late-type galaxies is instead dominated by the thermal
emission from dust. As in the Milky Way, the small dust grains emitting in the Mid-IR have an 
excess of emission if compared to big grains emitting in the Far-IR. While the Far-IR emission of galaxies increases
with the intensity of the interstellar radiation field, their Mid-IR emission is non--linearly 
related to the UV radiation field. The spectral energy distributions of the target galaxies
indicate that there is a linear relationship between the UV radiation field and the Mid-IR
emission of galaxies for low or intermediate activities of star formation, while the emission from
the hot dust seems to drop for strong UV fields. The Mid-IR colour of late-type galaxies
is not related to their activity of star formation.

The properties of the dust emission in the Mid-IR
seem more related to the mass than to the morphological type of the target galaxy.
Since the activity of star formation is anticorrelated to the mass of galaxies, this
reflects a relationship between the emission of dust in the Mid-IR and the
UV radiation field: galaxies with
the lowest Mid-IR emission for a given UV field are low mass, dwarf galaxies. These observational
evidences are easily explained if the carriers of the Unidentified Infrared Bands that dominate
the 6.75 $\mu$m emission are destroyed by the intense UV radiation field of dwarf 
galaxies, although abundance effects can also play a role.


\end{abstract}

\section{Introduction}

The bulk of the infrared (IR) emission of the interstellar medium (ISM) of 
galaxies is due to the thermal emission of the dust 
heated by the interstellar radiation field. Stellar photons  
are absorbed by the interstellar dust 
and re-emitted in the IR from $\sim$ 4 $\mu$m to $\sim$ 1 mm. 
The dust emitting in the Far-IR ($\lambda$ $\ga$ 70 $\mu$m) is probably composed of big 
grains (with a linear radius $r$ $\ga$ 200 \AA) of graphite and silicate in thermal equilibrium 
with the UV and optical photons of the interstellar radiation field
(D\'esert \etal ~\cite{Desert}; Dwek \etal ~\cite{D97}). 
In the range 10 $\mu$m $\la$ $\lambda$
$\la$ 70 $\mu$m the emission is usually dominated by very small, three dimensional 
grains (VSG) mostly composed of graphite
(10 \AA $\la$ $r$ $\la$ 200 \AA), absorbing mainly in the UV, whose emission is however
not negligible also at shorter wavelengths (D\'esert \etal ~\cite{Desert}; Dwek \etal 
~\cite{D97}).
In the range  3 $\mu$m $\la$ $\lambda$ $\la$ 15 $\mu$m the emission of the interstellar medium
is generally dominated by the Unidentified Infrared Emission Bands (UIB). The carriers of the UIBs and associated
continuum, often called Polycyclic Aromatic Hydrocarbons (PAHs) although this assignment
is uncertain, are heated  
stochastically by absorption of single photons and temporarily reach very high 
temperatures, at which most of the emission occurs (D\'esert \etal ~\cite{Desert}; Dwek \etal 
~\cite{D97}).

The knowledge of the dust emission properties of the ISM strongly increased in the last years
thanks to the large amount of data obtained by the IRAS satellite, 
which made an all--sky survey in 4 different filters, at 12, 25, 60 and 100 $\mu$m. 
More recently, the ISO satellite is providing higher quality data in the spectral
range $\sim$ 3 $\mu$m $<$ $\lambda$ $<$ 200 $\mu$m. In order
to analyse the properties of the IR emission of different classes of objects and/or 
of single galaxies, the ISOCAM and the ISOPHOT consortia constructed a coordinated program of
observations of an optically selected, complete sample of late-type galaxies in the Virgo cluster.
This sample, which includes a large number of galaxies of different morphological type and
luminosity, has been selected to be used as a reference for other studies. 
Furthermore the sample
includes objects in the centre and in the periphery of the cluster in order to study the effects
of the environment on the dust emission and on the evolution of galaxies. Because all 
the selected galaxies are at the same distance, the construction of a reference sample
is quite direct.

The selected sample has been observed entirely with CAM and partly with PHOT and LWS. 
In this paper we report on the integrated properties of the galaxies observed with CAM,
thus in the spectral range 3 $\mu$m $<$ $\lambda$ $<$ 20 $\mu$m. 
Preliminary results have been already published in Boselli \etal (\cite{BISO97}, Paper I)
and Boselli \& Lequeux (\cite{BL97}). The data, the analysis of the relationship between the star 
formation activity and the Mid-IR emission of the few resolved galaxies and the study of the
effects of the environment on the properties of the dust emission of late-type galaxies
will be discussed in future papers (Boselli \etal, in preparation).
The purpose of the present paper is to present a reference for other ISOCAM observations.
The sample is described in Sect. 2; a brief description
of the observations and of the data reduction is given in Sect. 3. In Sect. 4 we analyse the
Mid-IR luminosity distribution of the target galaxies, their spectral energy distribution (SED),
and the relationships between the Mid- and Far-IR emission of galaxies of different types and 
luminosities. The discussion and the conclusions are given in Sect. 5.

\section{The sample}

The selected sample is extensively described in Boselli \etal (\cite{B97}).
Briefly, the sample has been extracted from the Virgo Cluster Catalogue (VCC) of 
Binggeli \etal (\cite{VCC}) which is complete to B$_T$ = 18 mag. It includes all galaxies with
morphological type later than S0, with B$_T$ $\leq$ 18 mag, classified as cluster members
by Binggeli \etal (\cite{VCC}; \cite{Bing93}). Because of time limitations,
the sample has been restricted to include only galaxies within a projected distance of
2 degrees from M87, thus belonging to the core of the cluster, and comparison objects in the cluster periphery.  
The cluster periphery is at an angular distance larger than 4 degrees from the position of the maximum projected
galaxy density given by Sandage \etal (\cite{S85}), but excluding galaxies within 1.5 degrees
of the M49 sub--cluster. The total sample includes 117 objects with morphological types between
S0/a and Im and BCDs. Because of the low visibility of the Virgo cluster during
the ISO mission and because of the low detection rate of small
galaxies with ISOCAM, the 18 Im and BCDs in the cluster core could not be observed.
The resulting sample is still unbiased but includes only 
99 objects, with 26 spirals in the cluster core, and
73 spirals, irregulars and BCDs in the cluster periphery.
The observed sample thus constitutes an optically--selected, volume--limited complete sample,
ideal for statistical analyses.
A few early--type galaxies were serendipitously observed in some of the fields. These do 
not constitute a complete sample in any sense, but because of their random selection, 
they can be used as an unbiased reference sample of early--type objects.
Given the different densities of galaxies and of the 
intergalactic medium of the two selected regions, the observed sample can be used to study
the effects of the environment on the dust properties of galaxies. However we show in Sect. 4.1 
that the integrated Mid--IR fluxes of the spirals are not significantly affected by the
environment, so that the present sample is a true reference sample.

\section{Observations and data reduction}

A complete description of the observations and data reduction will be presented in a forthcoming
paper (Boselli et al., in preparation). The observations were part of the ISO Central
Program: all galaxies were observed during one of the few visibility windows of the
Virgo cluster during the ISO mission, in the summer of 1996.
The 32$\times$32 pixel long--wavelength camera was used to make a raster map
covering largely the galaxy, with a shift (then overlap) of 16 pixels
between successive positions. The pixel size was $6\arcsec\times6\arcsec$.
3$\times$3 rasters were made for galaxies with the smallest angular dimensions, while larger
rasters were adopted for giant or extended objects. 
The elementary integration time was 2 seconds, with 16 to 20 integrations per position.
To minimize the effects of transients of the detector on the flux calibration of the source,
40 to 70 integrations per position were used in the first raster position every time
the background was expected to change significantely from the previous target, thus
for significantly different pointings and for different filters.
In order to minimize the overhead time for galaxy pointing and for the stabilisation of the 
detector, galaxies were observed in a concatenated mode, with a complete sequence of observations
of 5-10 nearby galaxies in a given (and fixed) filter.
All galaxies were observed in 2 filters, LW2 ($\lambda$ = 6.75 $\mu$m, $\Delta$$\lambda$=5.00-
8.50 $\mu$m) and LW3 ($\lambda$ = 15.0 $\mu$m, $\Delta$$\lambda$=12.0-18.0 $\mu$m).
Images were reduced following the standard CIA/IDL\footnote{
CAM Interactive Analysis, CIA, is a joint development by the ESA Astrophysics Division,
and the ISOCAM Consortium led by the ISOCAM P.I., C. Cesarsky, Direction des Scienecs de la 
Mati\`ere, CEA, France}
 procedures available at the ISOCAM
center at the CEA in Saclay: images were dark--subtracted, flat-fielded, deglitched, corrected
for the transient response of the detector, and combined.
The total fluxes were determined using the IMCNTS routine in IRAF/Xray 
after subtraction of the sky background; counts were transformed in fluxes using
the conversion factors given in the ISOCAM manual. The absolute fluxes are believed to
be accurate to within 30\%, mainly due to calibration errors, but the relative fluxes
of the observed galaxies are certainly more accurate. 
The detection rate is 70\% at 6.75 $\mu$m and 55\% at 15.0 $\mu$m (see Table 1). Note that 
none of the Im galaxies have been detected. For undetected galaxies, an upper limit has
been determined as:

\begin{equation}
{F_{limit(\lambda)} = 2 \sigma_{\lambda} (5 \times 5 pixels)   \rm{(mJy)}}
\end{equation}

\noindent
where $\sigma$$_{\lambda}$ is the background noise per pixel at a given $\lambda$. We
thus assume that a galaxy has an extended emission (5 $\times$ 5 pixels, or 30 $\times$ 30
arcseconds).

\section{The Mid--IR emission of galaxies}

In this section we analyse the Mid--IR integrated properties of the observed galaxies
by comparing the CAM observations to data at other wavelengths: Far--IR data at 60 and
100 $\mu$m are taken from IRAS (Thuan \& Sauvage \cite{TS92}), near--IR magnitudes and 
optical data from
Boselli \etal (\cite{B97}) and Boselli (in preparation), and UV data
at 2000 \AA ~from Deharveng \etal (\cite{Deharveng}), Donas \etal 1998 (in preparation) 
and references therein.

\subsection{Mid--IR luminosity distribution}

Since the observed sample is complete, it can be used to determine the luminosity
distribution of galaxies of different types in the Mid--IR. The poor statistics for early--types 
and the low
detection rate for the dwarf irregulars (see Table 1), 
allow us to determine a reliable luminosity distribution only for spirals.

\begin{table}
\caption{Statistics and detection rate for the observed galaxies}
\label{Tab1}
\[
\begin{array}{p{0.15\linewidth}lllll}
\hline
\noalign{\smallskip}
Type   & observed  &  detected           &  & K' \\
       &           &  6.75 \mu m & 15 \mu m &    \\
\noalign{\smallskip}
\hline
\noalign{\smallskip}
E*             &    1   &   1 &  1 &  1  \\
S0*            &    2   &   2 &  2 &  2  \\
S0/a           &    5   &   5 &  3 &  5  \\
Sa             &   11   &  10 & 10 & 11  \\
Sab            &    4   &   4 &  4 &  4  \\
Sb             &    3   &   3 &  3 &  3  \\
Sbc            &    1   &   1 &  1 &  1  \\
Sc**           &   22   &  20 & 17 & 21  \\
Scd            &    4   &   4 &  2 &  4  \\
Sd-Pec-Amorph. &    6   &   4 &  3 &  6  \\
Sm             &    9   &   6 &  6 &  9  \\
Im             &   21   &   0 &  0 & 10  \\
BCD            &    9   &   5 &  3 &  6  \\
Sm/BCD         &    2   &   2 &  2 &  2  \\
Im/BCD         &    4   &   2 &  0 &  2  \\
\noalign{\smallskip}
\hline
\end{array}
\]
*: serendipitously observed in the field of nearby selected galaxies
**: one of the 22 Sc galaxies (VCC 1018),
serendipitously observed in the field of a nearby target, is
not a cluster member and will not be used in the following analysis.
\end{table}

The Mid--IR luminosity distribution at 6.75 and 15 $\mu$m is shown for different morphological 
classes in Fig. 1a and b. Sm and Amorphous galaxies are included in the spiral class, while the Sm/BCD
and the Im/BCD are included in the BCD class. The Mid--IR luminosities have been computed 
as:

\begin{equation}
{L_{\nu} = 4 \pi D^2 F_{\nu}\delta_{\nu}}
\end{equation}

\noindent
where $\delta$$_{\nu}$ is the filter width;
at 6.75 ($\delta$$_{6.75 \mu m}$ = 11.57 10$^{12}$ Hz) and 15 $\mu$m 
($\delta$$_{15 \mu m}$ = 5.04 10$^{12}$ Hz) $L_{\nu}$ is given by:

\begin{equation}
{L_{6.75 \mu m} = 4 \pi D^2 F_{6.75 \mu m} 2.724 \times 10^2  L_{\odot}}
\end{equation}

and

\begin{equation}
{L_{15 \mu m} = 4 \pi D^2 F_{15 \mu m} 1.184 \times 10^2  L_{\odot}}
\end{equation}

\noindent
where $D$ is the distance in Mpc (we adopt 17 Mpc for Virgo, as recently determined from
Cepheid distances observed with the HST; see for a recent review van den Bergh, 1996), 
and $F_{6.75 \mu m}$
and $F_{15 \mu m}$ are the fluxes at 6.75 and 15 $\mu$m, in mJy.

\begin{figure*}
\vbox{\null\vskip 18.3cm
\includegraphics{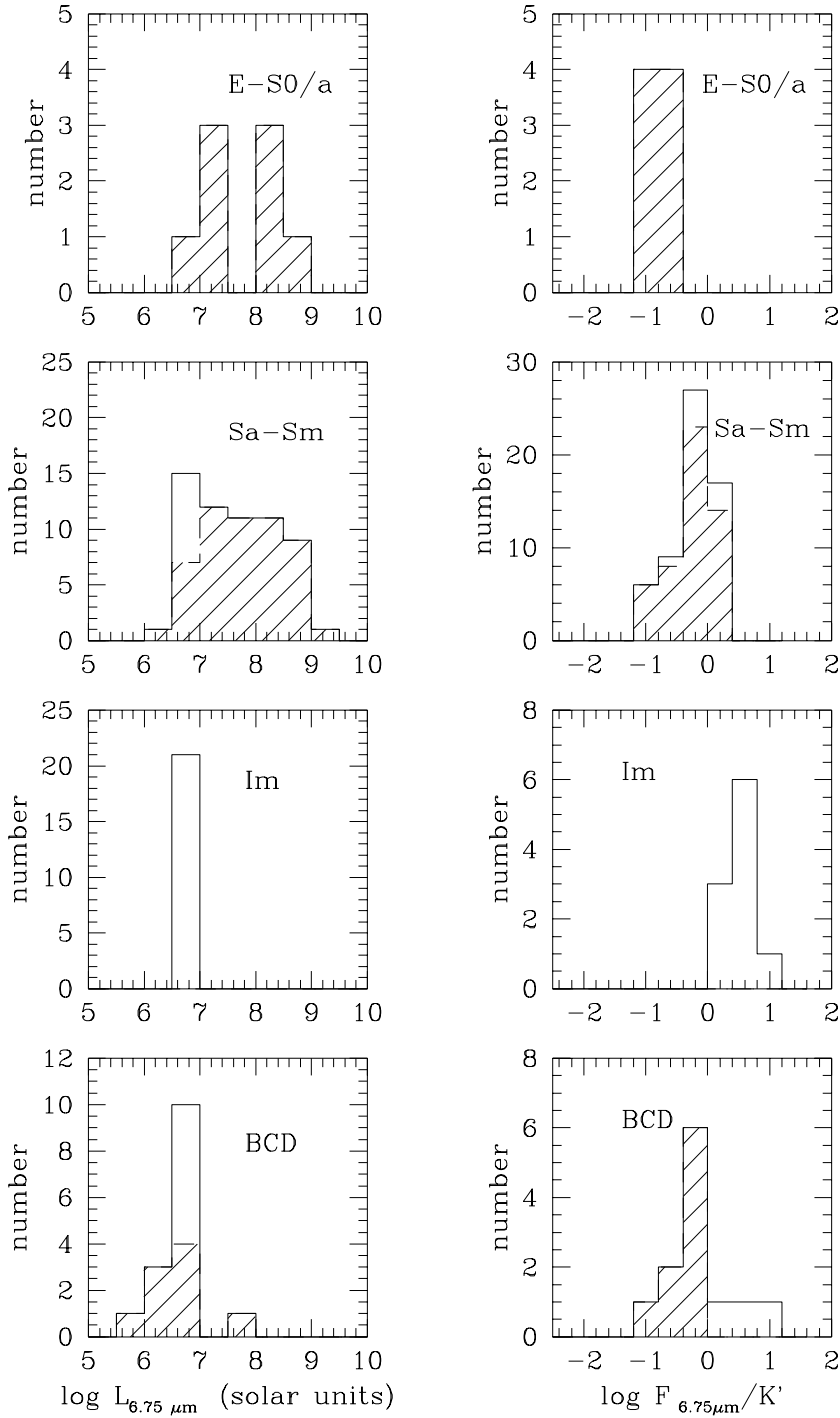}
}
\caption{The Mid--IR luminosity distribution at 6.75 $\mu$m (left column) and the one
normalized to the flux in the K' band (right column) for detected (dashed line, shaded histogram)
and undetected (full line, empty histogram) galaxies of different morphological type.
11 Ims and 3 BCDs with no K' magnitudes are excluded from the normalized luminosity distribution.} 
\label{Fig.1a}
\end{figure*}

\begin{figure*}
\vbox{\null\vskip 18.3cm
\includegraphics{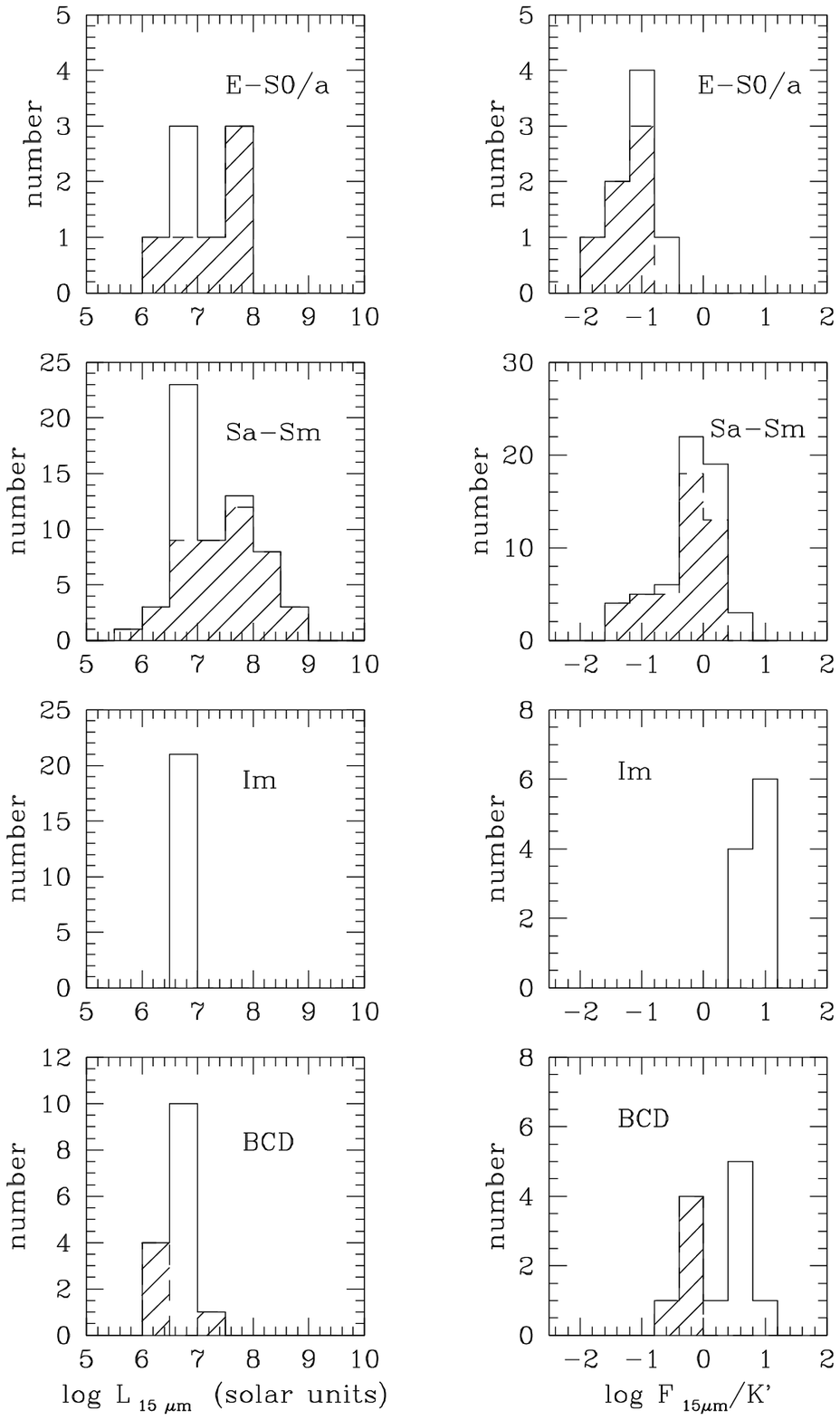}
}
\caption{The Mid--IR luminosity distribution at 15.0 $\mu$m (left column) and the one
normalized to the flux in the K' band (right column) for detected (dashed line, shaded histogram)
and undetected (full line, empty histogram) galaxies of different morphological type. 
11 Ims and 3 BCDs with no K' magnitudes are excluded from the normalized luminosity distribution.} 
\label{Fig.1b}
\end{figure*}

As shown by Sandage and collaborators (\cite{S85}), 
the optical luminosity function of spiral, elliptical and S0 galaxies in the Virgo cluster ends at B$_T$ = 16 mag,
while dwarfs reach lower magnitudes. Our sample being complete to B$_T$ = 18 mag,
since includes all the Virgo spirals in the limited observed volume. 90 \% and 82 \% of the
observed spirals (Sa-Sd-pec-Amorph.) are
detected at 6.75 and 15 $\mu$m respectively, the Mid--IR luminosity distribution
shown in Fig. 1 can be taken as representative at least
for the spiral population.
Spiral, elliptical and S0 galaxies have Mid--IR luminosities ranging from some 10$^6$ to 10$^9$
solar luminosities both at 6.75 and 15 $\mu$m, while dwarfs have Mid--IR luminosities generally
lower than 10$^7$ $L_{\odot}$.
In order to compare galaxies of different size and mass 
we construct the normalized Mid--IR luminosity
distribution by dividing the Mid--IR fluxes of the target galaxies by the flux in the near--IR 
K' band,
which is a good indicator of the total mass of galaxies (Gavazzi \etal~\cite{GPB96}, see Fig. 1a
and b).
The K' flux (in mJy) is defined as:

\begin{equation}
{F_{\rm{K'}} {\rm{(mJy)}} = 0.63 \times 10^6 \times 10^{{\rm{-K'magc}}/2.5}}
\end{equation}

\noindent
where K'magc is the K' magnitude corrected for extinction.
The K' magnitudes have been obtained by integrating along circular annuli the NICMOS 3 images 
of the target galaxies to the optical diameter at the 25 mag arcsec$^{-2}$ isophote, 
as described in Boselli et al. (1997b). As for the Mid--IR data, the K' magnitudes
are not extrapolated values; they represent the total emission of the galaxy.
As indicated in Table 1, this can be done for all the selected spirals and early-types,
but K' magnitudes are not available for a few Ims and BCDs with B $>$ 16.0 
(see Boselli \etal~\cite{B97}).

We checked the influence of the environment by comparing the luminosity distribution
of galaxies in the cluster core to that of galaxies in the periphery. Since dwarf galaxies
have been observed only in the periphery of the cluster (see Sect. 2), the present test was
performed only for spirals (Sa-Sm). A survival analysis (Isobe \etal~\cite{Isobe}), 
done to take into account upper limits, has shown, at least to the first order, 
no statistically significant differences between the normalized
Mid-IR emission of spirals in the core and in the periphery of the cluster.
This result can be deduced by the inspection of Fig. 2, where the log F$_{6.75\mu m}$/K'
and log F$_{15.0\mu m}$/K' are plotted versus the angular distance from the cluster centre
(for type Sa--Sm). 
Galaxies in the
core of the cluster have mean values of log F$_{6.75\mu m}$/K' = -0.35 $\pm$ 0.38 and
log F$_{15.0\mu m}$/K' = -0.52 $\pm$ 0.59, while galaxies in the periphery have 
log F$_{6.75\mu m}$/K' = -0.22 $\pm$ 0.37 and log F$_{15.0\mu m}$/K' = -0.27 $\pm$ 0.50.
The results of the analysis on the effects of the environment on the Mid-IR emission
of galaxies will be presented in a forthcoming paper (Boselli et al., in preparation).

To summarize, while the Mid--IR emission (per unit mass) of the spirals and later systems are comparable, 
the average 6.75 and 15 $\mu$m to K' flux ratio of E, S0 and S0/a galaxies is significantely 
lower than that of spirals. At 15 $\mu$m, where the difference is clearer, spirals have
on average a Mid--IR emission per unit mass higher by more than one order of
magnitude than E-S0/a.
BCD galaxies have, on average, a Mid--IR emission per unit mass comparable 
to that of spirals. If Im galaxies had a Mid--IR emission per unit mass 
comparable to that of spirals, their low luminosities would make them undetectable by ISOCAM.

\begin{figure}
\vbox{\null\vskip 13.0cm
\includegraphics{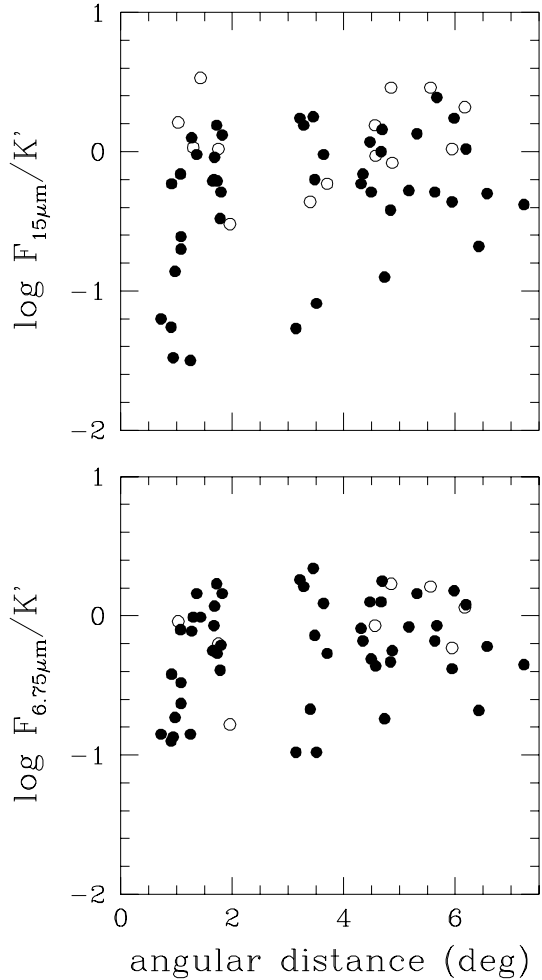}
}
\caption{The relationship between the log F$_{6.75\mu m}$/K' (down)
and log F$_{15.0\mu m}$/K' (up) and the angular distance from the cluster centre for
galaxies of type Sa--Sm. Galaxies in the core are all objects with an angular distance from
the cluster centre $\leq$ 2 degrees. Empty dots are for undetected objects.} 
\label{Fig.2}
\end{figure}

\subsection{The spectral energy distribution}

We constructed the spectral energy distribution (SED) from the
UV at 2000 \AA ~to the Far--IR at 100 $\mu$m for all the detected galaxies. 
They include, in order of increasing wavelength,
UV data at 2000 \AA, U (3650 \AA), B (4440 \AA), V (5500 \AA), J (1.25 $\mu$m), H
(1.65 $\mu$m) and K' (2.10 $\mu$m) photometry, 
6.75 $\mu$m (ISOCAM), 12 $\mu$m (IRAS), 15 $\mu$m (ISOCAM), 
60 and 100 $\mu$m (IRAS) data. Optical and near--IR magnitudes are values integrated
to the optical diameter at the 25 mag arcsec$^{-2}$ isophote, 
determined as described in Gavazzi \& Boselli (1996).
\footnote{For the few galaxies (5) with optical aperture or CCD photometry not available 
in the B band we used the photographic magnitudes from the Virgo Cluster Catalogue of
Binggeli \etal (\cite{VCC})}
Optical and near--IR data for spiral galaxies with type $<$ Sd
are corrected for galactic and internal extinction as described in Gavazzi \& Boselli (\cite{GB96}). As shown by Buat \& Xu (\cite{BX96}), the extinction for galaxies with
type $\geq$ Sd is $\sim$ 0.2 mag in the UV at 2000 \AA, thus only $\sim$ 0.1 mag at optical 
wavelengths and even smaller in the near-IR. For this reason we prefer to not apply any
internal extinction correction for Sd-Sm-Im and BCD galaxies.

Buat \& Xu 
(\cite{BX96}) have shown that at 2000 \AA ~the extinction is low ($\sim$ 0.9 mag),
even in bright spirals; this is confirmed by the lack of relationship between the 
UV/K' flux ratio and the inclination of the sample galaxies, that we have checked.
Because of the large uncertainty in the 
determination of the extinction in the UV, we prefer to adopt uncorrected values. 

In order to compare the SED of galaxies with different luminosities,
all fluxes are normalized to the near--IR K' flux and displayed in order of increasing
morphological type (see Fig. 3). To estimate the 
contribution of the cold stellar component to the Mid--IR emission of the target galaxies,
we plot for reference in Fig. 3 a black body at 3500 K adjusted on the peak of the 
emission of the cold stellar component, which is generally in the H band (dashed line).
 

\begin{figure*}
\vbox{\null\vskip 18.5cm
\includegraphics{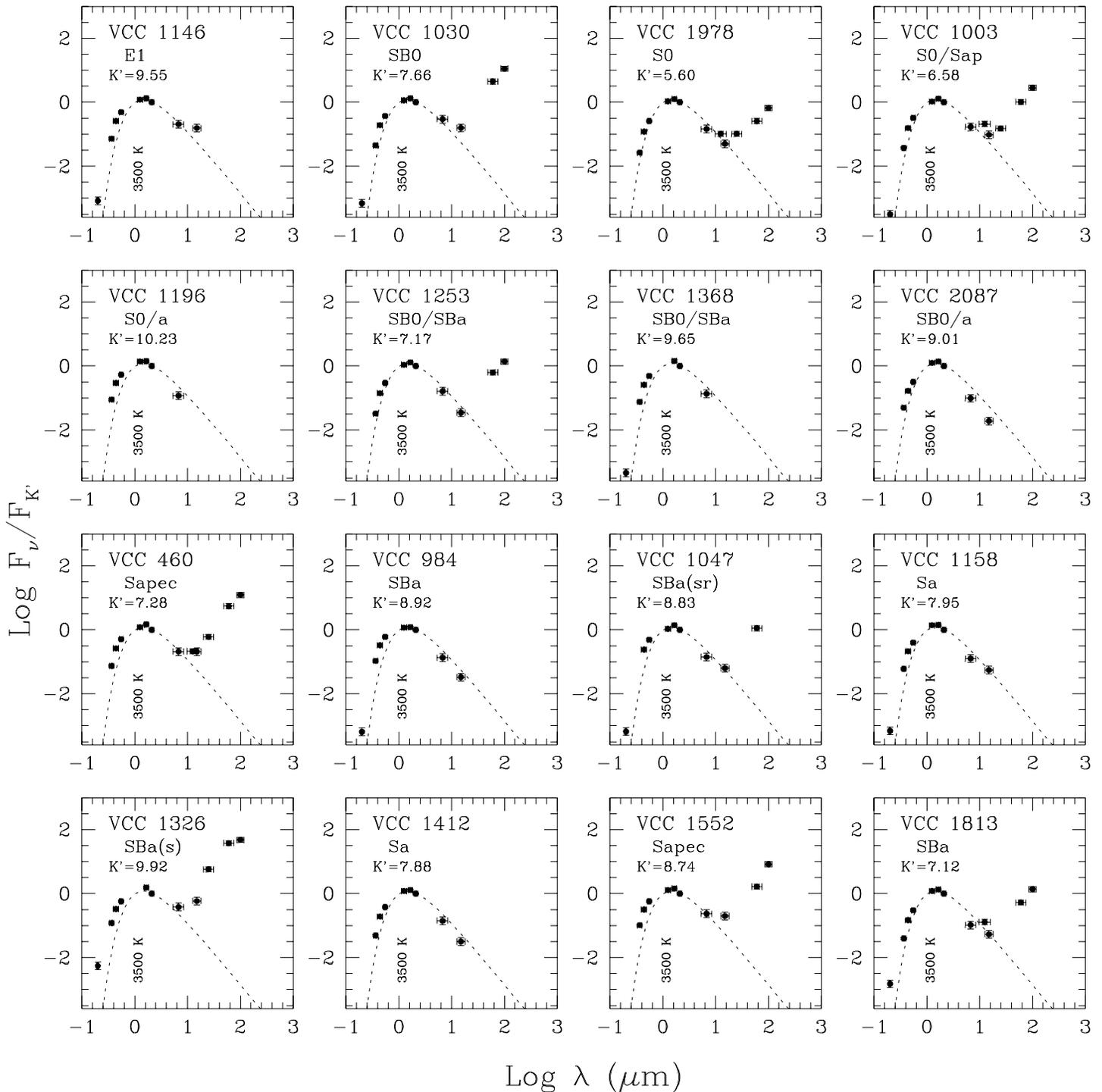}
}
\caption{The spectral energy distribution of the detected galaxies displayed in order of 
increasing morphological type. The dashed line indicates
a black body at 3500 Kelvin. The VCC name and the morphological type is given for each
galaxy.}
\label{Fig.3a}
\end{figure*}

\begin{figure*}
\vbox{\null\vskip 18.5cm
\includegraphics{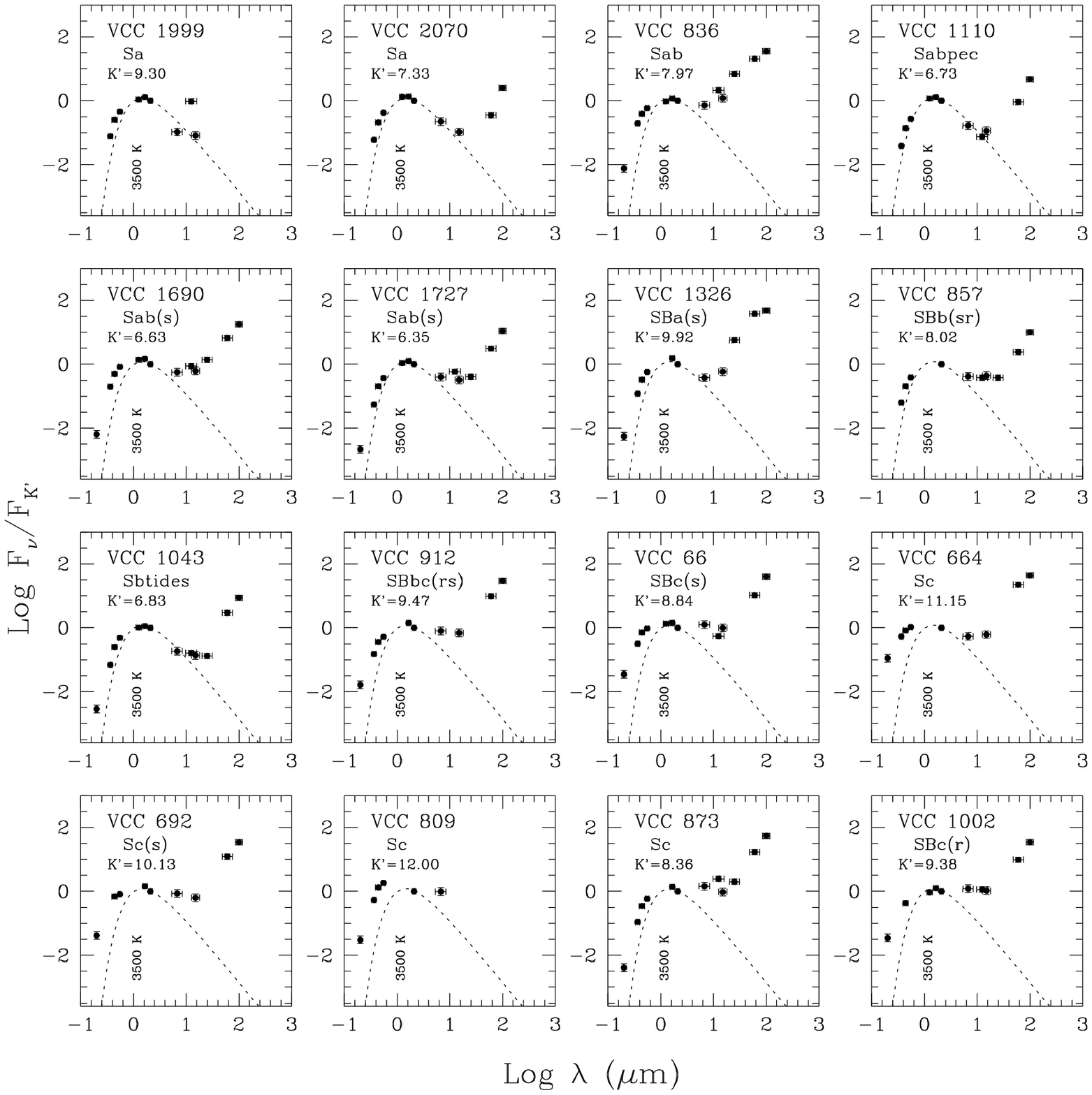}
}
\caption{Continued.}
\label{Fig.3b}
\end{figure*}

\begin{figure*}
\vbox{\null\vskip 18.5cm
\includegraphics{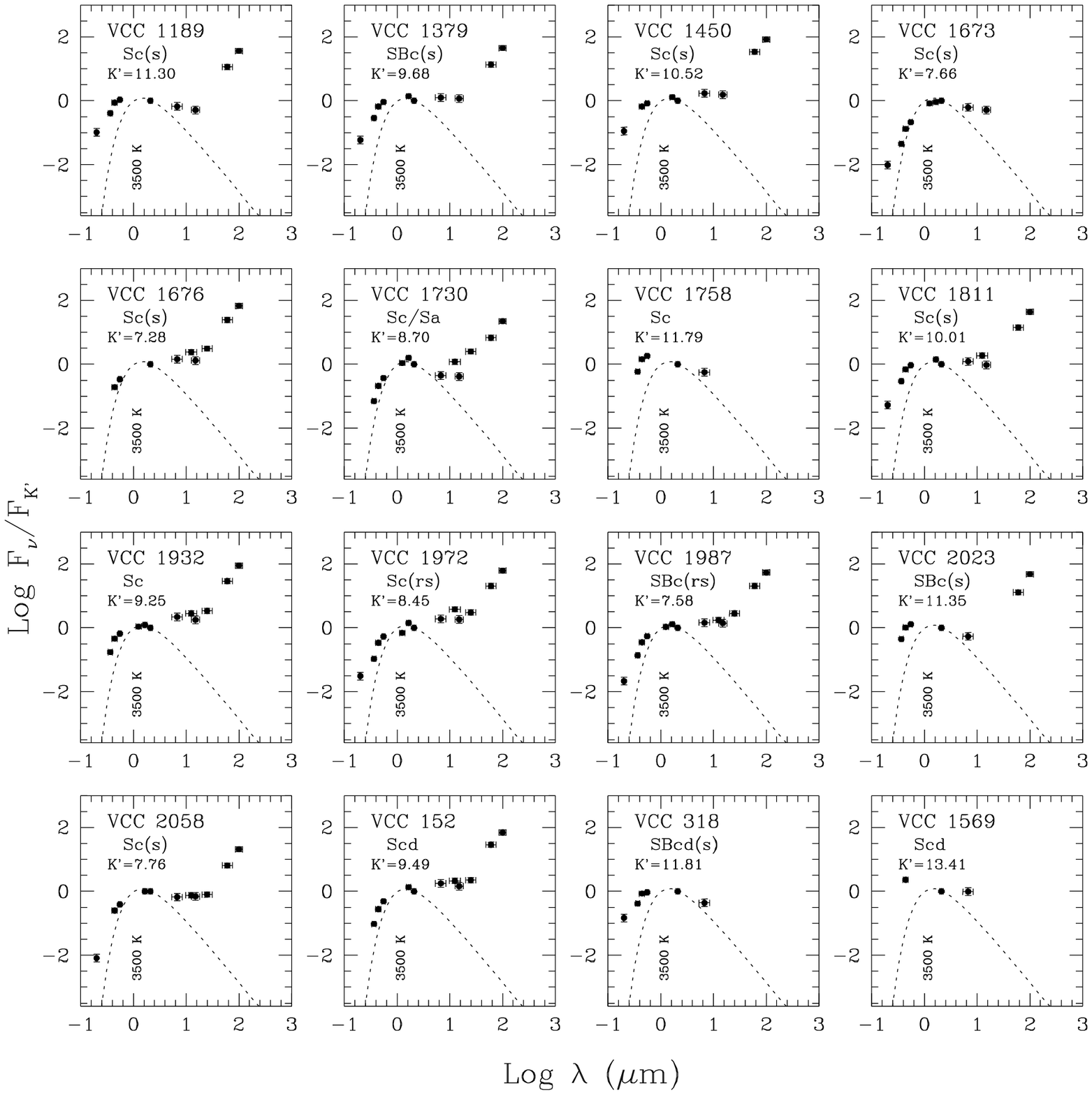}
}
\caption{Continued.}
\label{Fig.3c}
\end{figure*}

\begin{figure*}
\vbox{\null\vskip 18.5cm
\includegraphics{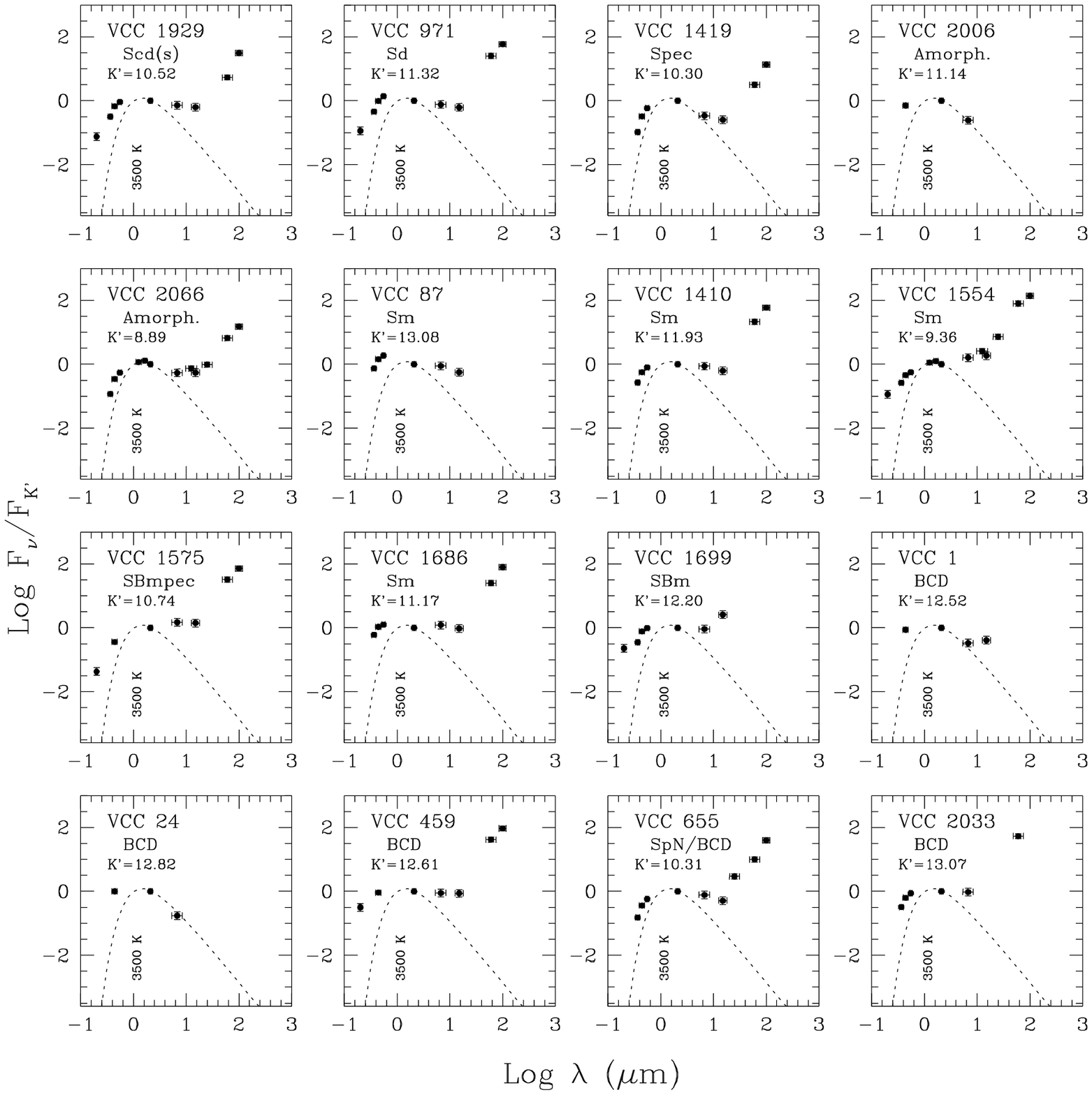}
}
\caption{Continued.}
\label{Fig.3d}
\end{figure*}

\begin{figure*}
\vbox{\null\vskip 5.5cm
\includegraphics{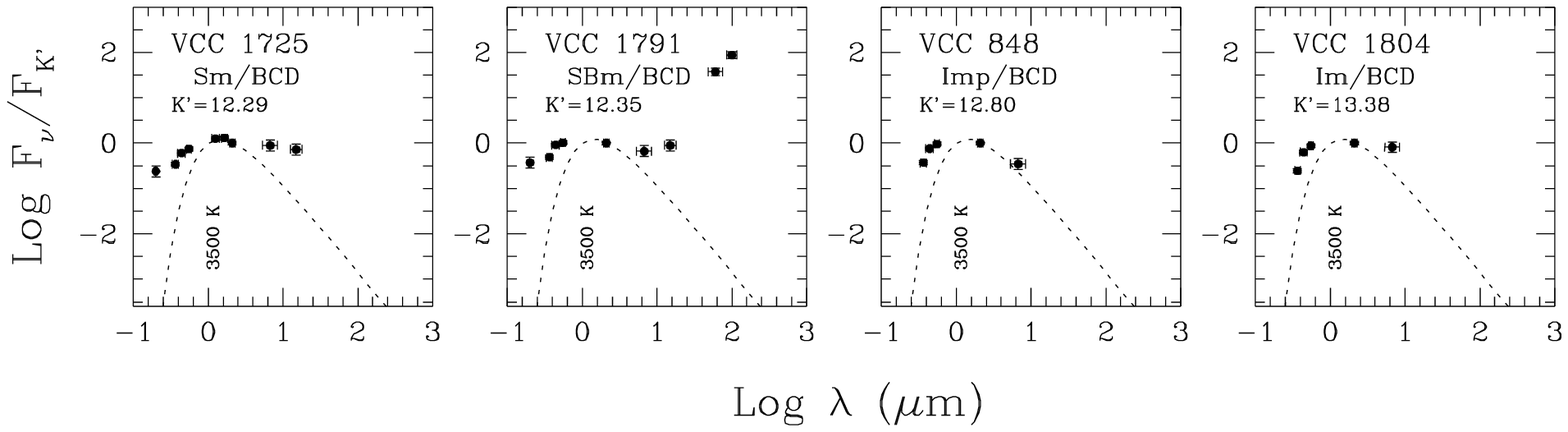}
}
\caption{Continued.}
\label{Fig.3e}
\end{figure*}

As described in Gavazzi \etal (\cite{GPB96}), the peak of the stellar emission is in the near--IR
for massive, early--type spirals, while it moves to shorter wavelengths in late--type, low 
luminosity systems. The UV emission is only important in the latter objects (Gavazzi \etal ~\cite{GPB96}).
The Mid--IR emission of early--type spirals can be partly due to the
Rayleigh-Jeans tail of the cold stellar component at least at 6.75 $\mu$m. 
The Far--IR emission of spirals
and dwarfs seems correlated with the presence of young stars emitting in the UV. The Scds and 
the Sms have the highest Mid--IR 
emission per unit mass, while lower--luminosity Ims and BCDs, which are characterised by a
stronger UV radiation field, have on average a 6.75 and 15 $\mu$m 
emission comparable to that of Sc galaxies. 
In all spirals and dwarfs the flux density at 6.75 $\mu$m is comparable to that at 15 $\mu$m.
In the few early--type objects (Ellipticals, S0 and S0/a) the flux density at 15 $\mu$m is 
significantely lower than that at 6.75 $\mu$m and seems to follow the Rayleigh-Jeans tail
of the cold stellar component.
From this we conclude that the Mid--IR
emission is dominated by the emission of the stars, as also shown by
the analysis of the Mid--IR/Far--IR colour-colour relation
(Boselli \& Lequeux ~\cite{BL97}; Madden \etal ~\cite{M97}).

In Table 2 we list the average values (in logarithmic scales) of the 6.75 $\mu$m to K',
15 $\mu$m to K' and 6.75 to 15 $\mu$m flux ratios for the early-type objects in our sample.

\begin{table*}
\caption{Mid--IR to near--IR properties of the early-type galaxies of our sample; 
values are given in logarithmic scales}
\label{Tab2}
\[
\begin{array}{p{0.15\linewidth}llllll}
\hline
\noalign{\smallskip}
VCC   & NGC  &  type & F_{6.75\mu m}/K' & F_{15\mu m}/K' &  F_{6.75\mu m}/F_{15\mu m} \\
\noalign{\smallskip}
\hline
\noalign{\smallskip}
1146  &  4458  &  E1        &   -0.69     &      -0.81    &       0.13   \\
1030  &  4435  &  SB0       &   -0.53     &      -0.80    &       0.28   \\
1978  &  4649  &  S0        &   -0.84     &      -1.30    &       0.46   \\
1003  &  4429  &  S0/Sap    &   -0.77     &      -1.02    &       0.25   \\
1196  &  4468  &  S0/a      &   -0.93     &      <-0.72   &       >-0.21 \\
1253  &  4477  &  SB0/SBa   &   -0.79     &      -1.46    &       0.66   \\
1368  &  4497  &  SB0/SBa   &   -0.87     &      <-0.98   &       > 0.11 \\
2087  &  4733  &  SB0/a     &   -1.01     &      -1.72    &       0.71   \\
\noalign{\smallskip}
average (detected galaxies only)&        & &  -0.80     &      -1.18    & 0.42         \\
standard deviation:   &                  & &   0.15     &       0.37    & 0.22         \\ 
\noalign{\smallskip}
\hline
\end{array}
\]
\end{table*}

The average value found for the early--type
sample of Virgo galaxies listed in Table 2 is similar to that determined for the S0 
VCC 1978 (NGC 4649), whose 6.75 $\mu$m diffuse emission, contrary to spiral 
galaxies, is not associated with any event of star formation as traced by the H$\alpha$ image,
as shown by Boselli \& Lequeux (\cite{BL97}).
This observational evidence, along with the fact that most of the SEDs of the 
Mid--IR emission of the early--type 
galaxies in the present sample seem to follow the Rayleigh-Jeans tail of the 
cold stellar component at about 3500 K, suggests that the average value
listed in Table 2 can be used, at least as a first approximation, to empirically estimate
the stellar contribution to the Mid--IR emission of later--type galaxies.

Boselli \etal (\cite{BISO97}) have shown that there is a clear correlation between
the Far--IR emission and the UV flux of galaxies, while in the Mid--IR the 6.75 and 15 $\mu$m
normalized fluxes are proportional to the star formation rate only for low or moderate 
activities. Galaxies with strong UV radiation fields have, in general, a lower Mid--IR
emission. This result can be seen in Fig. 4, where we plot the distribution of the spectral flux (a) and
energy (b; defined as $\nu$F$_{\nu}$ normalized to the energy in the K' band) for galaxies 
divided into 4 different classes of UV/K' flux ratios. Arbitrary thresholds on the UV/K'
ratios are chosen to include in each group a similar number of objects.
The difference between the UV to K' flux ratio of galaxies belonging to the 4 different classes
is not the result of a selection due to the internal extinction in the UV flux, since all galaxies
are well mixed in inclination, but is real and reflects the properties of the UV radiation field
in their ISM.

\begin{figure*}
\vbox{\null\vskip 19.0cm
\includegraphics{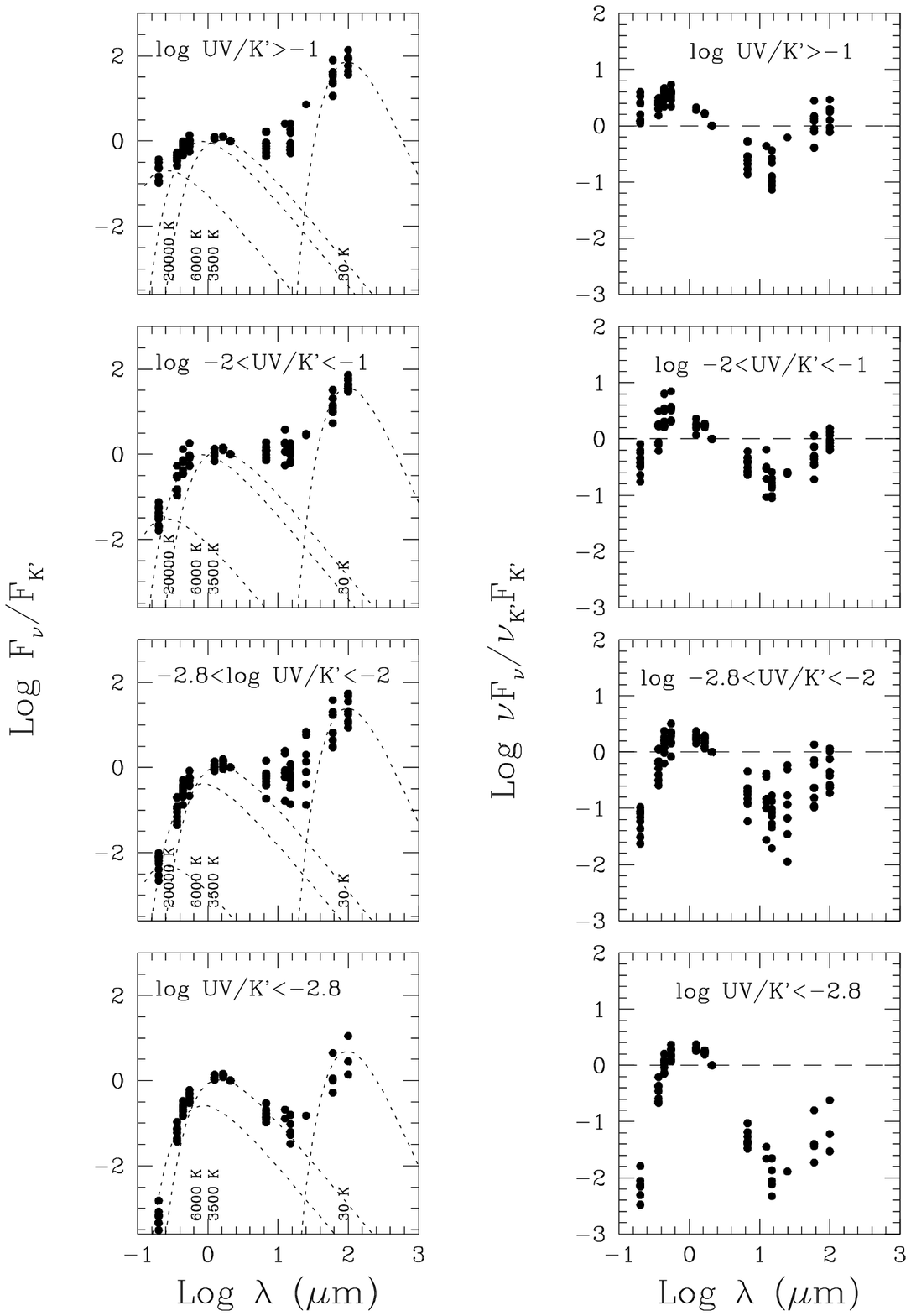}
}
\caption{The spectral energy distribution in fluxes (a) and in energy (b) defined as $\nu$F$_{\nu}$ of the detected galaxies in 4 different classes
of UV/K' flux ratios. The dashed line indicate 
black bodies at 20000, 6000 and 3500 degree Kelvin and a modified black body
($\lambda$$^{-2}$) for the dust component at 30 K.} 
\label{Fig.4}
\end{figure*}

Black bodies with a temperature of T=20000 K, 6000 K and 3500 K are fitted by eye to the UV, optical
and near-IR data, while a modified ($\lambda$$^{-2}$) 
black body with T=30 K is fitted by eye to the cold dust component
emitting in the Far--IR. 
The stellar emission of galaxies with log UV/K' $<$ -2.8 is dominated by the cold component
peaked in the near--IR. A hot stellar component (T=20000 K) is not needed to explain the
UV emission of these galaxies, which is probably due to stars of intermediate mass and 
temperature.
The cold stellar component ($\lambda$ $\sim$ 1 $\mu$m) also dominates 
the peak of the energy distribution, with an energy balance $\sim$ 100:1 between the optical-near-IR and the UV at 2000 \AA. A typical case is the S0/Sapec VCC 1003. 
The Mid-IR emission of these galaxies is dominated by the Rayleigh-Jeans tail of the 
cold stellar component at about 3500 K. 
As for galaxies with log UV/K' $<$ -2.8, the stellar emission and the energy distribution 
of galaxies with -2.8 $<$ log UV/K' $<$ -2.0 is dominated by the cold component
peaked in the near--IR. A typical case is the Sab(s) VCC 1727.
These galaxies are quiescent in the Mid--IR and in the Far--IR if
compared to the intermediate--class objects (-2.0 $<$ log UV/K' $<$ -1.0), which are also 
characterised by a higher contribution of the young stellar component to their global stellar 
emission (as for example VCC 1987). 
The galaxies with the most intense UV radiation fields (log UV/K' $>$ -1.0), whose stellar
emission and energy distribution are peaked at optical wavelengths, have the most 
intense Far--IR emission for the 
sample galaxies. However their Mid--IR emission is comparable to that of the
objects with an intermediate star formation activity. A typical case is the BCD galaxy VCC 459.
It is also interesting to mention that in some BCDs (VCC 1, VCC 24, VCC 848) 
the emission at 6.75 $\mu$m follows the SED of a cold stellar component, while at
15 $\mu$m the dust emission is important. However if the stellar emission of these galaxies
is dominated by hot stars, a black body at 3500 K overestimates the stellar flux at 6.75 $\mu$m,
and there must be some dust emission at this wavelength.

To identify the heating sources of the dust emitting in the Mid-- and Far--IR, we plot on Fig.
5 the relationships between the 6.75, 15 and 100 $\mu$m fluxes normalized to the K' flux
(first line) and to the UV flux (second line) versus the UV to K' flux ratio, which is 
widely used as a star formation tracer. 

\begin{figure*}
\vbox{\null\vskip 9.5cm
\includegraphics{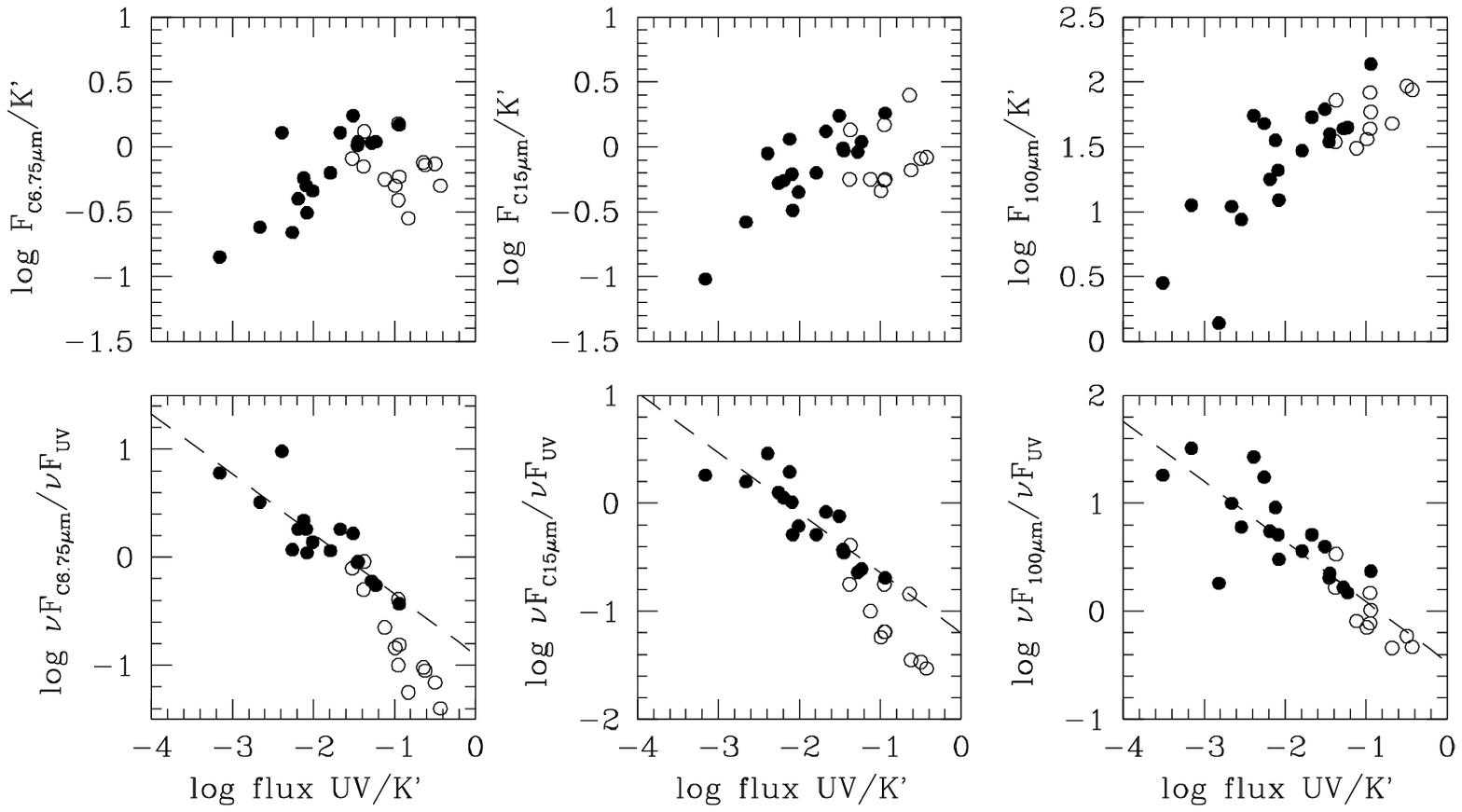}
}
\caption{First line: the relationship between log F$_{C6.75\mu m}$/K', log F$_{C15.0\mu m}$/K' 
and log F$_{100\mu m}$/K' and the UV flux (normalized to the K' flux); second line:
the relationship between log $\nu$$_{6.75\mu m}$F$_{C6.75\mu m}$/$\nu$$_{2000 \AA}$UV, log $\nu$$_{15.0\mu m}$F$_{C15.0\mu m}$/$\nu$$_{2000 \AA}$UV 
and log $\nu$$_{100\mu m}$F$_{100\mu m}$/$\nu$$_{2000 \AA}$UV and the 
UV flux (normalized to the K' flux). The Mid--IR fluxes are
corrected for the stellar contribution as described in Sect. 4.3.
Open symbols indicate galaxies with K' $>$ 10 mag, filled
dots are for objects with K' $<$ 10 mag. The dashed line is the best fit to the
log $\nu$$_{100\mu m}$F$_{100\mu m}$/$\nu$$_{2000 \AA}$UV vs. log UV/K' relationship, shifted 
in the log $\nu$$_{C6.75\mu m}$F$_{C6.75\mu m}$/$\nu$$_{2000 \AA}$UV and log $\nu$$_{C15.0\mu m}$F$_{C15.0\mu m}$/$\nu$$_{2000 \AA}$UV relationships to follow the data for bright galaxies (filled dots)} 
\label{Fig.5}
\end{figure*}

The strong relationship observed between the flux at 100 $\mu$m and the UV flux 
confirms that the big grains responsable for the 100 $\mu$m emission are heated by UV photons.
The ratio of the energies emitted in the Far--IR and in the UV is 
strongly related to the UV flux, is close to unity in
low-mass galaxies (open dots), and is $\ge$ 10 in high-mass, quiescent early-type spirals
(filled dots). This strong 
relationship can be partly due to the effect of the extinction, which is expected to be higher
in higher-mass, quiescent galaxies (Buat \& Xu ~\cite{BX96}), and in part to the fact that optical photons might significantely contribute to the heating of the dust grains in the quiescent sample. 
As shown by Boselli \etal (\cite{BISO97}), in low-mass galaxies with a strong star formation
activity (open dots) the Mid-IR emission at 6.75 $\mu$m and marginally at 15 $\mu$m, is anticorrelated with the intensity of the UV radiation field, while the fluxes at 6.75 and 15
$\mu$m are well correlated with the UV flux in massive spirals (filled dots). 
At the same time, the energy balance between UV photons and the Mid--IR emission does not hold 
in galaxies with an intense UV radiation field because these galaxies are relatively
transparent to UV radiation.
The Far--IR to UV energy ratio versus UV/K' relationship is shared in the Mid--IR only
by massive spiral galaxies; in the dwarf population (open dots) the relationship between 
$\nu$$_{6.75\mu m}$F$_{C6.75\mu m}$/$\nu$$_{2000 \AA}$UV, 
$\nu$$_{15.0\mu m}$F$_{C15.0\mu m}$/$\nu$$_{2000 \AA}$UV and the UV to K' flux ratio 
has steeper slopes, indicating that the relation between the inciding UV photons and
the emitting dust grains changes in strong UV radiation fields. 
The contribution of optical photons to the heating of the dust grains
emitting in the Mid--IR is negligible because of their small
physical dimensions, thus the Mid--IR to UV energy ratio vs. UV/K'
relationships are probably entirely due to the extinction.

\subsection{Mid--IR and Far--IR dust emission of late--type galaxies}

\begin{figure*}
\vbox{\null\vskip 15.0cm
\includegraphics{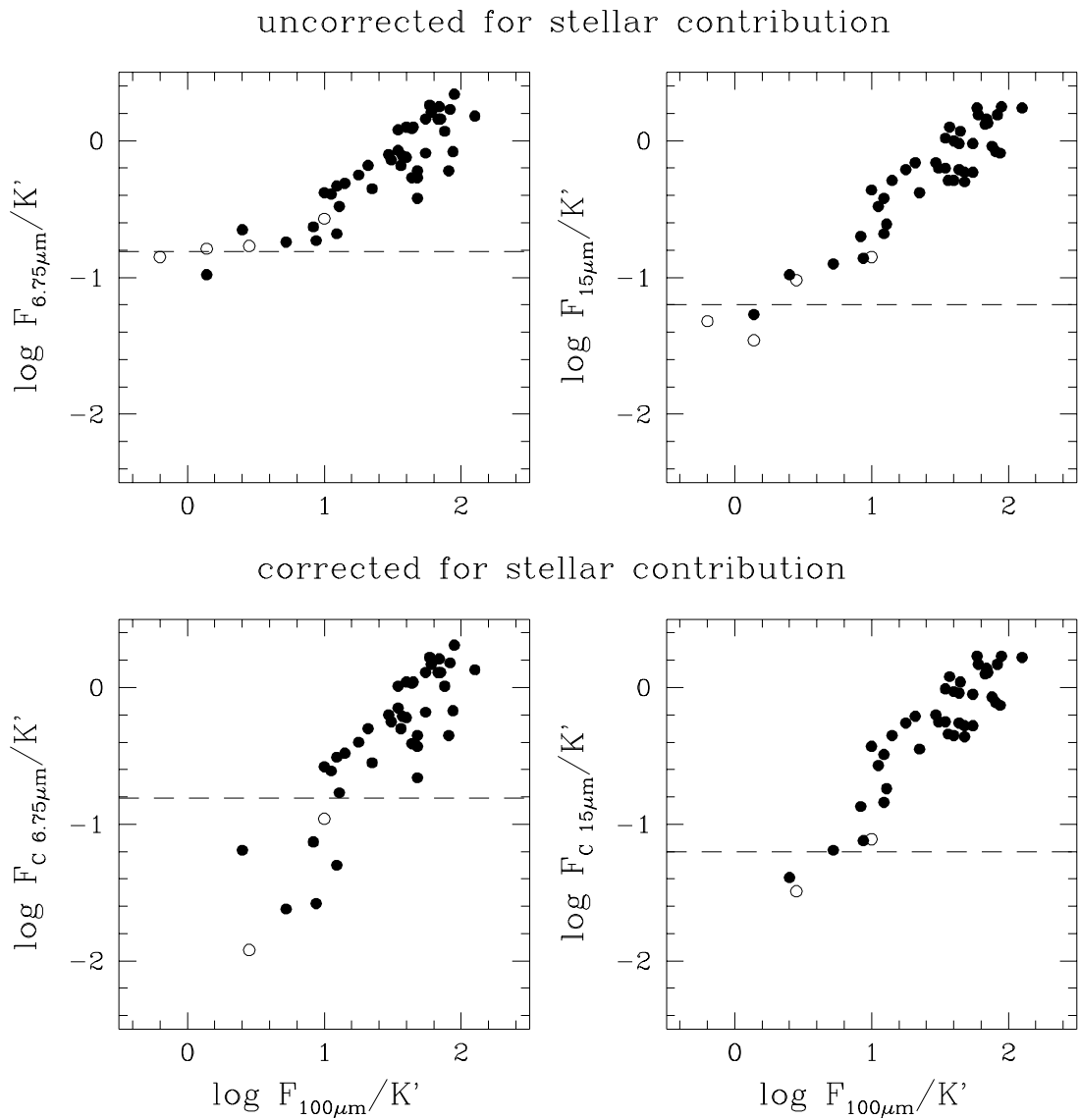}
}
\caption{The relationship between the normalized Far--IR emission at 100 $\mu$m  and the Mid--IR 6.75 $\mu$m and 15 $\mu$m emission
uncorrected (upper panels) and corrected (lower panels) for stellar contribution
for the detected galaxies. Open symbols indicate early--type
galaxies, filled symbols late--type objects. The dashed
line indicate the average value found for early--type galaxies.} 
\label{Fig.6}
\end{figure*}

Using as template the small sample of early--type galaxies
listed in Table 2, we can estimate and remove the average stellar contribution to the total
Mid--IR emission of the target galaxies. 
We thus obtain that the corrected fluxes at 6.75 and 15 $\mu$m are given by the relations:

\begin{equation}
{Log F_{C6.75 \mu m}/K' = Log (F_{6.75 \mu m}/K' - 0.158)}
\end{equation}

\noindent
and

\begin{equation}
{Log F_{C15.0 \mu m}/K' = Log (F_{15.0 \mu m}/K' - 0.063)}
\end{equation}

As shown in Fig. 6, when the corrected Mid--IR
values are used, a relationship between the Mid--IR and
the Far--IR emission is observed even at low dust emissions.
Given the large uncertainty in the determination of the stellar contribution to the Mid--IR
emission of early--type galaxies, this method is reliable only for objects whose emission at
6.75 and 15 $\mu$m is dominated by dust, thus we exclude from the following analysis 
all objects with log F$_{6.75\mu m}$/K' $<$ -0.55.
Corrected fluxes at 6.75 and 15 $\mu$m are used throughout the following analyses.
In Fig. 7 we plot the Mid-- and Far--IR colours as function of the UV/K' flux ratio, the K' magnitude
and the morphological type, coded as in Gavazzi \& Boselli (\cite{GB96}):
3=Sa, 4=Sab, 5=Sb, 6=Sbc, 7=Sc, 8=Scd, 9=Sd, amorphous or peculiar, 10=Sm, 11=Im, 12=BCD, 13=Sm/BCD, 14=Im/BCD.

\begin{figure*}
\vbox{\null\vskip 20.0cm
\includegraphics{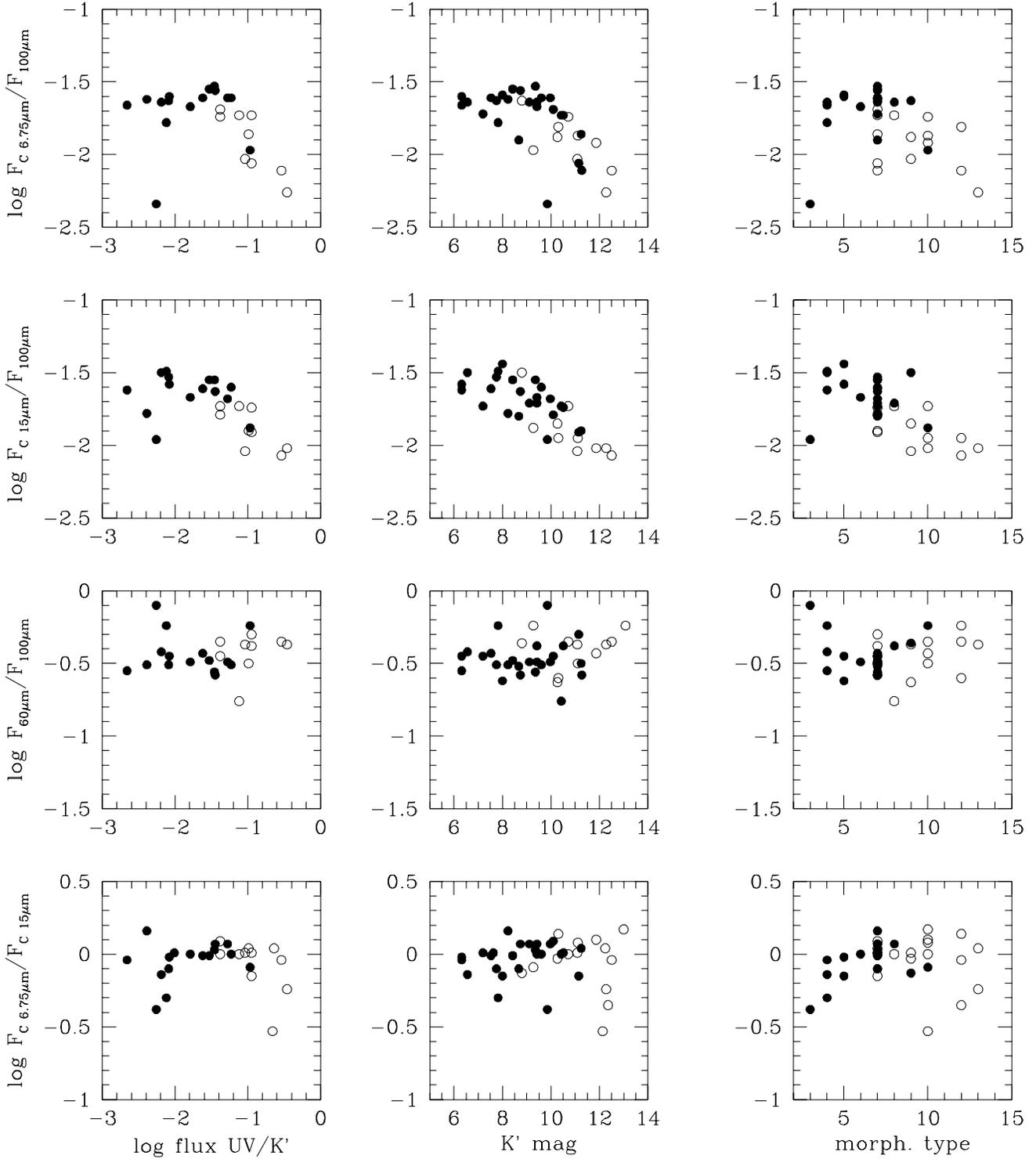}
}
\caption{The relationship between the Mid-- and Far--IR colours F$_{C 6.75\mu m}$/F$_{100\mu m}$ (first line),
F$_{C 15\mu m}$/F$_{100\mu m}$ (second line), F$_{60\mu m}$/F$_{100\mu m}$ (third line) 
and F$_{C 6.75\mu m}$/F$_{C 15\mu m}$ (fourth line) and the normalized UV flux (first column; open dots are galaxies with K' mag $>$ 10, filled dots objects with K' mag $<$ 10), the K' magnitude
(second column; open dots are galaxies with morphological type later than Scd, filled dots
with morphological type earlier than Sd) and the morphological type coded as 
described in the text (third column; open dots are galaxies with K' mag $>$ 10, filled dots objects with K' mag $<$ 10). The Mid--IR flux densities have been corrected for stellar contribution as described in the text; all galaxies have uncorrected log F$_{6.75\mu m}$/K' $>$ -0.55.
The object with F$_{C 6.75\mu m}$/F$_{100\mu m}$=-2.35 and F$_{C 15\mu m}$/F$_{100\mu m}$
which does not follow the general trends is the Sa galaxy VCC 1326. The SED
of this object shows that at least at 6.75 $\mu$m the Mid-IR flux can be strongly contaminated
by stellar emission.} 
\label{Fig.7}
\end{figure*}

Figure 7 clearly shows that the dust properties in the Mid-- and in the Far--IR are
completly different in galaxies of different morphological type and/or mass or luminosity,
characterised by a different UV interstellar radiation field.
While the Far--IR F$_{60\mu m}$/F$_{100\mu m}$ and Mid--IR F$_{C 6.75\mu m}$/F$_{C 15\mu m}$ flux ratios are completely
independent of the UV radiation field, the mass and the morphological type of the parent
galaxy, the mixed F$_{C 6.75\mu m}$/F$_{100\mu m}$ and F$_{C 15\mu m}$/F$_{100\mu m}$ flux ratios are strongly anticorrelated
with the UV radiation field, the K' luminosity and the morphological type.
\footnote{The object with F$_{C 6.75\mu m}$/F$_{100\mu m}$=-2.35 and F$_{C 15\mu m}$/F$_{100\mu m}$
which does not follow the general trend in Fig. 7 is the Sa galaxy VCC 1326. The SED
of this object shows that at least at 6.75 $\mu$m the Mid-IR flux can be strongly contaminated
by stellar emission even if its log F$_{6.75\mu m}$/K' $>$ -0.55.}
First of all it must be noticed that both the Far-- and the Mid--IR colours F$_{60\mu m}$/F$_{100\mu m}$ and
F$_{C 6.75\mu m}$/F$_{C 15\mu m}$ are quite similar in galaxies spanning a large range (more than 2 orders of
magnitude) of star formation rates, and thus that at least for "normal", optically selected galaxies the infrared colours cannot
be used as tracers of star formation. The same result has been obtained using other less
direct star formation indicators such as the B-K' and U-B colours, and is valid for galaxies 
of all morphological types later than Sa, spanning a large range in absolute magnitude.
This result seems in contrast with the general assumption that Far--IR colours are good
indicators of star formation (Helou \cite{Helou}). It must be however noticed that most of
the previous work was based on Far--IR selected samples which includes Far--IR bright
galaxies such as starburst and active galaxies (Helou \cite{Helou}), not included in the Virgo
optically selected sample. Sauvage \& Thuan (\cite{ST94}) have however shown that the 60 to
100 $\mu$m flux ratio does not increase monotonically from E-S0 to late-type spirals and 
irregulars as expected from current dust models (e.g. D\'esert \etal ~\cite{Desert})
if the Far--IR colour is simply
related to the UV radiation field or to the star formation activity of the parent galaxy.
Sauvage \& Thuan (\cite{ST94}) interpreted this observational evidence as the result of the fact
that, at least in galaxies of type E-S0 to Sbc, the Far--IR colours are mostly controlled
by the spatial distribution of dust relative to stars, and not by the star formation activity
as in later-type objects. Given the large intrinsic dispersion in the Far-IR colours in each
morphological class, the present data do not exclude that the trend observed by 
Sauvage \& Thuan (\cite{ST94}) is present.

As already shown by Boselli \etal (\cite{BISO97}), the mixed colours F$_{C 6.75\mu m}$/F$_{100 \mu m}$ and F$_{C 15\mu m}$/F$_{100\mu m}$ are anticorrelated with the UV radiation field: galaxies with a strong
star formation activity have in general a higher Far--IR to Mid--IR emission. 
Given the strong inverse relationship between the surface density of young ionizing stars and 
the luminosity or mass of a galaxy (Gavazzi \etal ~\cite{GPB96}), 
the strong anticorrelation between the flux ratios F$_{C 6.75\mu m}$/F$_{100 \mu m}$ and F$_{C 15\mu m}$/F$_{100\mu m}$ and the UV flux 
can be translated into a well defined relationship between the flux ratios F$_{C 6.75\mu m}$/F$_{100 \mu m}$ and F$_{C 15\mu m}$/F$_{100\mu m}$ and the mass of a galaxy. A relationship with a larger dispersion is also
observed between the mixed Mid-- to Far--IR colours and the morphological type.
It is interesting to note that it is the near--IR luminosity more than the morphological
type that discriminates the behaviour of the Mid-- to Far--IR flux ratio of a given object, with 
luminous and massive galaxies characterized by a given F$_{C 6.75\mu m}$/F$_{100 \mu m}$ and F$_{C 15\mu m}$/F$_{100\mu m}$
flux ratio and low mass objects with smaller Mid-- to Far--IR flux ratios.

\section{Discussion and conclusion}

The availability of an unbiased, optically--selected and complete sample 
allows us to statistically draw the average Mid--IR properties of "normal" galaxies.
We present for the first time the Mid--IR luminosity distribution at 6.75 and 15 $\mu$m
of a complete, optically-selected sample of late--type galaxies.

The present analysis clearly shows that the Mid--IR emission of ellipticals
and lenticulars (morphological range E--S0/a) is dominated by the Rayleigh-Jeans tail
of the photospheric emission of the cold stellar component emitting in the near--IR. This
is still true for some early--type quiescent spirals. Figure 3 however does not exclude
that some of these early--type objects have dust emitting in the Mid--IR. Madden \etal (\cite{M97}),
by analysing a large sample of early--type galaxies observed with ISOCAM, 
have in fact shown that many objects characterized by the presence of atomic or molecular gas,
or by nuclear star formation, have a Mid--IR emission already dominated 
by the dust at $\lambda$ $>$ 6 $\mu$m (Madden \etal (\cite{M97}); Madden et al. in 
preparation).

In late-type galaxies the Mid-IR emission is generally dominated by dust. The
strong relationship between the integrated UV flux at 2000 \AA ~and the Mid-IR
emission (both normalized to the mass of the galaxy) observed by Boselli \etal (\cite{BISO97})
indicates that the very small dust grains and the UIB carriers responsible for the emission
at 6.75 and 15 $\mu$m are heated preferentially by hot, massive stars. This
is also confirmed by the good morphological correspondance between the peaks
of the emission at 6.75 and 15 $\mu$m and the H$\alpha$ images of some
nearby, bright well--resolved late--type galaxies (Sauvage et al. 1996, Boselli \& Lequeux 1997).
However these dust grains and UIB carriers can also be heated by the visible light of evolved stars
if its radiation density is large,
as shown for example by the fact that elliptical galaxies with little UV
radiation or the bulge of M 31 also exhibit 6.75 and 15 $\mu$m dust emission
(Madden et al. in preparation; Cesarsky D. et al. in preparation; Uchida et al., 1997).

However in low-luminosity late--type galaxies Boselli et al. (1997a) have shown
that the integrated Mid--IR emission does not follow the
mean UV radiation field: the Mid-IR emission of these objects 
does not increase, and may even decrease for high
Far-UV fluxes. Figure 7 also shows that for such objects the ratio of the Mid-IR
flux to the Far--IR flux decreases at low K' luminosities or small total masses 
(and strong Far--UV radiation fields). All this points to a lack of very small
grains and UIB carriers in low-luminosity late--type galaxies, 
both per unit mass of gas and relative to the big
grains that are responsible for the Far--IR emission. There are two 
possible explanations for this trend. 

One can be the effect of metallicity (Sauvage et al. 1990). 
One expects a lower dust/gas
mass ratio for these galaxies since the grains are mostly made of heavy
elements, but this does not necessarily imply a lower small grain/large
grain ratio. However such a lower ratio could result from a second--order
effect: the lower abundance of carbon with respect to oxygen (and probably
to silicon) might lower the abundance of carbonaceous grains which are often
believed to dominate the Mid--IR emission in our Galaxy, relative to that
of silicate grains which are less efficient emitters (see e.g. Dwek et al.
1997). Then the overall Mid--IR emission will decrease with respect to the 
Far--IR one which is believed to be always dominated by silicates. 

The other possible interpretation of the low Mid--IR/Far--IR emission ratio
has been proposed by Boulanger et al. (1988) who suggested that in radiation 
fields larger than $\sim$ 50 times that in the solar neighborhood up to
80 \% of the grains emitting in the Mid--IR can be destroyed by UV photons. 
Such a high radiation field exists in large fractions of the volume of the
dwarf galaxies with K' $>$ 10 mag.

With the present data we cannot chose between these two interpretations, which 
in fact may hold simultaneously. A study of the ISO data base when available
will probably allow to solve this problem.

\acknowledgements
 
We wish to thank V. Buat, A. Contursi, J.M. Deharveng and G. Gavazzi for interesting
discussions during the preparation of this work. A.B. thanks H. Facques for his hospitality
during his staying in Paris. We thank J. Donas for providing us unpublished UV data.
We thank the referee, B.T. Soifer, for interesting comments and suggestions.
\newline


\vskip 1.5cm


\begin{thebibliography}{}


\bibitem[1985]{VCC}
Binggeli B., Sandage A., Tammann G., 1985, AJ 90, 1681

\bibitem[1993]{Bing93}
Binggeli B., Popescu C., Tammann G., 1993, A\&AS 98, 275


\bibitem[1997]{BL97}
Boselli A., Lequeux J., 1997, in: "Extragalactic Astronomy in the Infrared",
Editions Fronti\`eres, ed. T. Thuan and G. Mamon, p237

\bibitem[1997b]{B97}
Boselli A., Tuffs R., Gavazzi G., Hippelein H., Pierini D., 1997b,
A\&AS, 121, 507

\bibitem[1997a]{BISO97}
Boselli A., Lequeux J., Contursi A., et al, 1997a, A\&A, 324, L13

 
\bibitem[1988]{Boul88}
Boulanger F., Beichman C., D\'esert F., et al., 1988, ApJ, 332, 328


\bibitem[1996]{BX96}
Buat V., Xu C., 1996, A\&A, 306, 61



\bibitem[1994]{Deharveng}
Deharveng J.M., Sasseen T., Buat V., \etal
1994, A\&A 289, 715

\bibitem[1990]{Desert}
D\'esert F.--X., Boulanger F., Puget J.--L. 1990, A\&A 237, 215



\bibitem[1997]{D97}
Dwek E., Arendt R., Fixsen D., et al., 1997, ApJ, 475, 565

\bibitem[1996]{GB96}
Gavazzi G., Boselli A., 1996, Astrophys. Letters and Comm. 35, 1

\bibitem[1996]{GPB96}
Gavazzi G., Pierini D., Boselli A., 1996, A\&A 312, 397



\bibitem[1986]{Helou}
Helou G. 1986, ApJ 311, L33

\bibitem[1986]{Isobe} 
Isobe T., Feigelson E.D., Nelson P.I., 1986, ApJ 306, 490


\bibitem[1997]{M97}
Madden S., Vigroux L., Sauvage M., 1997, in: "Extragalactic Astronomy in the Infrared",
Editions Fronti\`eres, ed. T. Thuan and G. Mamon, p229 


\bibitem[1985]{S85}
Sandage A., Binggeli B., Tammann G., 1985, AJ, 90, 1759


\bibitem[1990]{STV90}
Sauvage M, Thuan T., Vigroux L., 1990, A\&A, 237, 296

\bibitem[1996]{S96}
Sauvage M., Blommaert J., Boulanger F., et al., 1996, A\&A, 315, L89

\bibitem[1994]{ST94}
Sauvage M., Thuan T., 1994, ApJ 429, 153

\bibitem[1992]{TS92}
Thuan T., Sauvage M., 1992, A\&AS 92, 749

\bibitem[]{}
Uchida K., Sellgren K., Werner M., 1998, ApJ 493, L109

\bibitem[]{}
van den Bergh S., 1996, PASP, 108, 1091



\end{thebibliography}
\end{document}